\documentclass[twocolumn,prl]{revtex4-1}
\usepackage{amsmath,amssymb}
\usepackage{graphicx}
\usepackage{color}

\usepackage{times}

\begin{document}

\author{Maria Daghofer}

\affiliation{Institute for Theoretical Solid State Physics, IFW Dresden, 01171 Dresden, Germany}

\author{Masudul Haque}

\affiliation{Max-Planck-Institut f\"{u}r Physik komplexer Systeme, N\"{o}thnitzer Stra{\ss}e 38, D-01187 Dresden, Germany}

\date{\today}

\title{Viewpoint: Toward Fractional Quantum Hall physics with cold atoms}

\begin{abstract} 
  Viewpoint on 
  \begin{itemize}
  \item {\bf Reaching Fractional Quantum Hall States with Optical Flux Lattices}\\
    Nigel R. Cooper and Jean Dalibard, 
    Phys. Rev. Lett. {\bf 110}, 185301 (2013). 
  \item {\bf Realizing Fractional Chern Insulators in Dipolar Spin Systems}\\
    N. Y. Yao, A. V. Gorshkov, C. R. Laumann, A. M. Läuchli, J. Ye,
    and M. D. Lukin, 
    Phys. Rev. Lett. {\bf 110}, 185302 (2013). 
  \end{itemize} 
 \emph{Researchers propose new ways to recreate fractional quantum Hall physics using ultracold atoms
   and molecules.  }

\end{abstract}

\maketitle

In the fractional quantum Hall (FQH) effect,  
observed in two-dimensional electron gases in a
magnetic field, the Hall resistance is quantized to 
non-integer multiples of $h/e^2$.  
The fractional values are not an accidental small deviation from the previously observed
\emph{integer} quantum Hall (IQH) effect, but instead point to a type of order that had not been
known previously, namely ``topological order''. One consequence of
topological order are 
fractionalized excitations, e.g., carrying fractional charge $e/3$, similar to the Hall resistance
of the FQH.  The idea of topological order with its unconventional properties has since been
generalized (e.g. to spin liquids), however, no further experimental realization 
has been unequivocally established.  Now, two papers by Cooper and
Dalibard~\cite{Cooper} and by Yao {\it et al.}~\cite{Yao} 
propose to reproduce FQH states with cold atoms and molecules.  The proposals differ from earlier
ones by building on recent theoretical research indicating alternative ways to mimic the impact of a
magnetic field~\cite{flatbandChern_2011_threePRLs}.  Since cold atomic and molecular gases enable experimental tunability far beyond what is
possible in solid-state setups, one can hope to custom build a variety of FQH-like states.

The experimental signature of the FQH and IQH effects
are quite similar --- step functions in conductance caused by
quantized transport in two-dimensional electron gases subject to a
magnetic field.  In both effects, magnetic fields split
the electron energy levels into Landau levels, each consisting of a
large number of degenerate single-particle states. The IQH states arise
when there are just enough electrons to fill an integer 
number of Landau levels, while the FQH states appear at fractional
fillings.   
In both effects, topological aspects of the electronic wave functions
are important, as the magnetic field also modifies the eigenstates. The Hall
conductivity of a filled Landau level turns out to be given by the (first) Chern number, a mathematical
concept from the field of topology. This realization allowed the prediction and later
realization of topological insulators (TI's), and thus a generalization of IQH physics beyond Landau
levels~\cite{topol_insul}. FQH states are topological in an even
stronger sense %-
\cite{Wen_topol_order_1990}: Their ground-state degeneracy depends on the topology of the
two-dimensional real-space surface they live on, e.g., it is the same for a
sphere and a cuboid, but different for a torus.

Topological order 
violates the long-held belief that ordering requires symmetry breaking. 
In a magnet, e.g., spins are aligned along a common direction chosen
spontaneously  from originally equivalent ones, thus breaking a symmetry. The local magnetization
measures how well a spin is aligned and thereby the strength 
of the order. Topological order, in contrast, does not break any symmetries  and has 
no local order parameter equivalent to the magnetization, it may 
thus at first sight not appear ordered at all. Instead, it
is ``non-local'' and encoded in the wave function of the whole system. Local perturbations
can consequently not destroy it, which makes such states appealing
for fault-tolerant quantum computation. 

\begin{figure}
\includegraphics[width=\columnwidth]{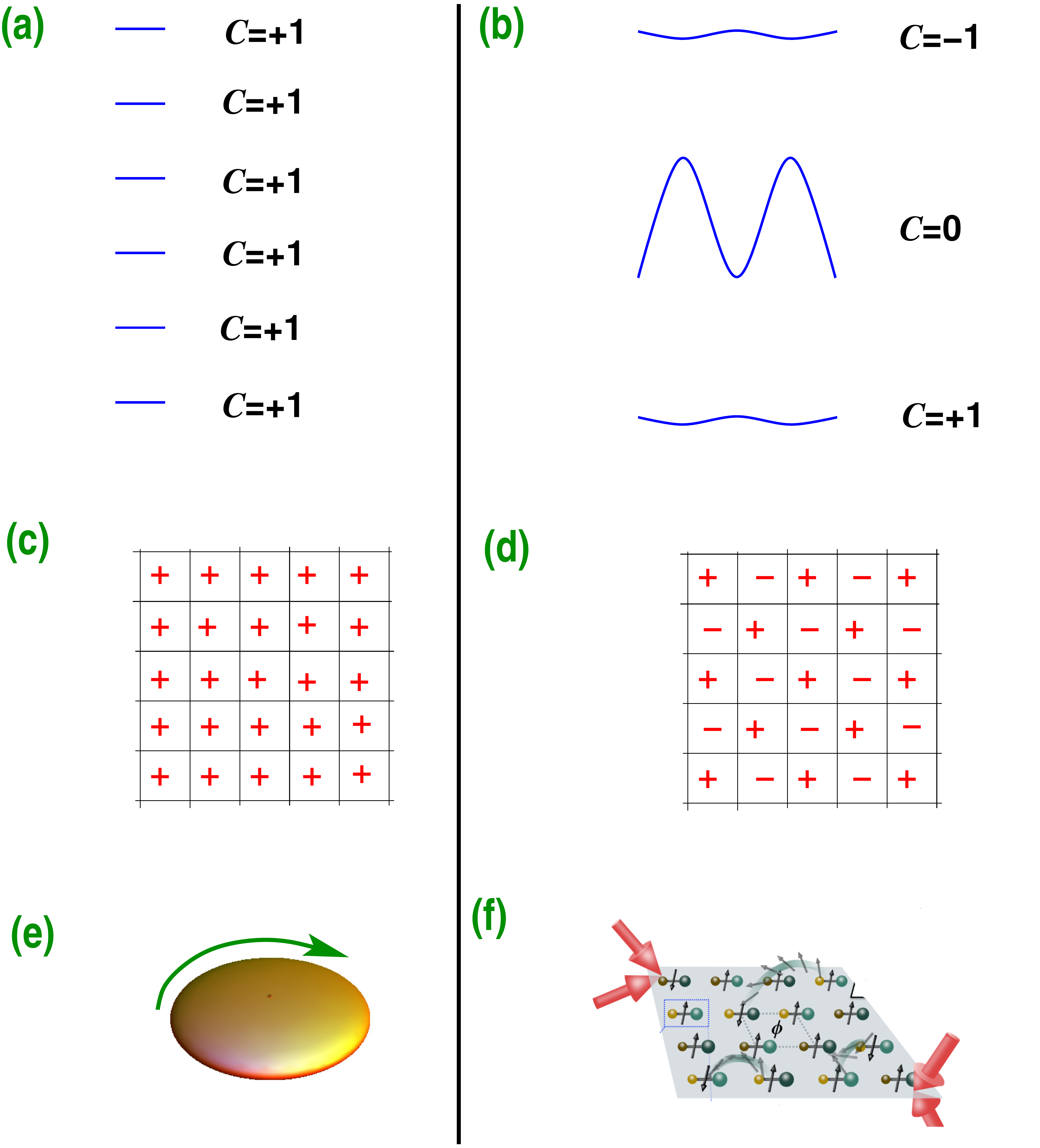}
\caption{ \label{figA} 
Contrast between the ``Landau level'' mechanism of realizing FQH states (left column) with ``Chern
insulator'' mechanism (right column).  (a) and (b) show examples of band structures.  The
traditional mechanism involves Landau levels, each with a Hall conductivity (given by the Chern
number $C$) of the same sign (a).  In Chern-band systems, the bands have different Chern numbers
(b).  In real space, an effective magnetic flux threads each elementary plaquette (i.e., each little
square). In the lattice version of Landau levels, each flux has the
same sign (c), just as the effective magnetic field in a
trapped rotating Bose gas (e)  has the same sign everywhere.
In Chern insulators, fluxes often alternate (d), as through a scheme like the one proposed in
Ref.~\cite{Yao}  using polar molecules in optical lattices (f).
}
\end{figure}

Unsurprisingly, the goal of realizing FQH physics and topological
order in ultracold atoms serving as ``quantum simulators''
has attracted interest and effort early on. 
Proposals involved rotating trapped bosonic
gases~\cite{Cooper_rotating_review}, creating effective magnetic
fields for lattice bosons~\cite{lattice_artificial_FQH} or exploiting
the internal structure of atomic states to create artificial gauge
fields~\cite{artificial_gauge}. Yet, despite intense research, the
goal of establishing FQH states in cold gases has remained elusive. 

The two papers by Cooper and Dalibard~\cite{Cooper} and by Yao {\it et al.}~\cite{Yao}   represent
a new generation of such proposals.  Instead of relying on Landau levels, they build on the
above-mentioned generalization of IQH physics 
to TI's. Even though the Hall
conductivity was introduced in the context of magnetic fields, 
it can be calculated for 
any filled band in a solid, yielding the 
Chern number $C$. It vanishes in most
cases, but a band where it is non-zero is said to be ``topologically nontrivial'' and a
spin-polarized band with $C\neq 0$ is referred to as a ``Chern band''. (Similarities and
differences between a lattice supporting Landau levels and one supporting Chern bands are sketched
in Fig. 1.) Accordingly, it was proposed three years ago that Chern bands may provide an alternative
route to FQH-like states~\cite{flatbandChern_2011_threePRLs} and substantial interest has gone in the
direction of these ``fractional Chern insulators'' (FCI's). Yet, despite a few materials-based
proposals~\cite{material_proposals}, an experimental realization has not been found so far.

In Ref.~\cite{Yao} , the authors propose to use polar
molecules to realize a bosonic FCI. In this proposal, molecules are
strongly bound to their place, but have a rotational degree of freedom. A
change in the rotational eigenstate can be seen as a spin flip from
the ``spin up'' ground state, which can in turn be interpreted as a
boson. Molecules at neighboring sites can exchange their rotational
eigenstates, which allows the boson to hop from site to site. The
crucial ingredient is now that the wave function picks up a phase
during the hopping process.  The phases associated with each bond are
distributed in such a way that a boson moving around one of the square
plaquettes shown in Fig. 1(d) acquires a phase of $\textrm{e}^{\pm
  i\phi}$, where the fluxes $\pm \phi$ alternate between
plaquettes. This is typical of Chern systems, and the two bands here
have indeed Chern numbers $C=\pm 1$, similar to Fig. 1 (b). 

The proposal in Ref.~\cite{Cooper} also proposes a
realization of an FCI system, but is based on earlier
ideas by Cooper and collaborators to formulate topological nontrivial
models in momentum space. The authors propose to create a periodic
spatial modulation of the coupling between lasers and bosonic atoms
(such as Rb atoms). Such a modulation creates phases for the
hoppings, which in turn again establish the nontrivial band topology
in a similar way as a strong magnetic field would. 

Since FQH states are driven by interactions between
particles, these should be large compared to the kinetic
energy. On the other hand, they should be
smaller than the gaps between bands, because a
mixture of states with $C=1$ and $C=-1$ might cancel to topologically
trivial $C=0$. Fulfilling both criteria is easiest for nearly flat
Chern bands, analogous to the high degeneracy of Landau levels. 
The tunability of cold gases permits this and the optimized band
structure given in Ref.~\cite{Yao}  has a ``flatness
ratio'' (band width divided by gap to the next 
band) of $f=11.5$, which has been found flat enough for FCI states in
other models. 
The authors consider interacting bosons
in this band and find a variety of phases.  The FCI
competes with superfluid and solid phases, but occupies a sizable
region in parameter space.  
In Ref.~\cite{Cooper} the ratio is even
better ($f=46$). By interpolating between this Chern band and a Landau level, the authors
provide numerical evidence for several different FCI states at different densities.

Compared to earlier proposals to find FQH states in cold quantum
gases, these two do not so much propose technological advances but
rather extend the potential routes to FQH-like states. Building on recent research suggesting that it may
not be necessary to copy all features of Landau levels, they propose
to keep only some aspects, namely
the Chern number of the band of interest and its reduced dispersion,
and leave out others (the Chern numbers of other bands, a constant ``magnetic
field''). The advantage over materials-based approaches to FCI
states is the flexibility of cold gases, which makes it 
appear more realistic to get into the needed parameter regimes.
An experimental realization of an FCI, especially a
highly tunable one, would allow to study topological order in far
greater depth than the original FQH setting and as a first step would
establish whether the proposed generalizations from Landau levels to Chern
bands hold.

\end{document}